\documentclass[aps,superscriptaddress]{revtex4}
\usepackage{amssymb,amsmath,latexsym,amsfonts,graphicx}
\usepackage{subfigure}
\usepackage[colorlinks=true, pdfstartview=FitV, linkcolor=blue, citecolor=red, urlcolor=magenta, breaklinks=true]{hyperref}
\begin{document}
\title{Absorption and scattering of a noncommutative black hole}

\author{M. A. Anacleto}\email{anacleto@df.ufcg.edu.br}
\affiliation{Departamento de F\'{\i}sica, Universidade Federal de Campina Grande
Caixa Postal 10071, 58429-900 Campina Grande, Para\'{\i}ba, Brazil}

\author{F. A. Brito}\email{fabrito@df.ufcg.edu.br}
\affiliation{Departamento de F\'{\i}sica, Universidade Federal de Campina Grande
Caixa Postal 10071, 58429-900 Campina Grande, Para\'{\i}ba, Brazil}
\affiliation{Departamento de F\'isica, Universidade Federal da Para\'iba, 
Caixa Postal 5008, 58051-970 Jo\~ao Pessoa, Para\'iba, Brazil}
 
\author{J. A. V. Campos}\email{joseandrecampos@gmail.com}
\affiliation{Departamento de F\'isica, Universidade Federal da Para\'iba, 
Caixa Postal 5008, 58051-970 Jo\~ao Pessoa, Para\'iba, Brazil}

\author{E. Passos}\email{passos@df.ufcg.edu.br}
\affiliation{Departamento de F\'{\i}sica, Universidade Federal de Campina Grande
Caixa Postal 10071, 58429-900 Campina Grande, Para\'{\i}ba, Brazil}

\begin{abstract} 
In this paper we focus on the partial wave method with the aim of exploring the scattering of massless scalar waves due to the noncommutative Schwarzschild black hole via Lorentzian smeared mass distribution. 
We determine the phase shift analytically in low-frequency limit and we show that the scattering and absorption cross section is modified. Specially, we show that in the low-frequency limit the scattering/absorption cross section has its value decreased when we increase the value of the non-commutativity parameter. In addition, we have confirmed this result by solving the problem numerically for arbitrary frequencies.
Such modifications found for the scattering and absorption cross section present similarities with the Reissner-Nordstr\"{o}m black hole.
\end{abstract}

\maketitle
\pretolerance10000

\section{Introduction}
Noncommutative spacetime in gravity theories have been the object of several investigations in the recent literature ---
see~\cite{Nicolini:2008aj,Szabo:2006wx} for comprehensive reviews. 
In particular, the study considering the effects of noncommutativity on black hole physics has been an area of great interest, mainly because of  the possibility of a better understanding of the final stage of the black hole due to its evaporation. 
As is well known the  noncommutativity eliminates point-like structures in favor of smeared objects in
flat spacetime~\cite{Smailagic:2003yb,Smailagic:2003rp}. 
Furthermore, one has been shown in~\cite{Nicolini:2005vd} that noncommutativity can be implemented in General Relativity by modifying the matter source.
{Therefore, noncommutativity is introduced by modifying mass density so that { the Dirac delta function is replaced by a Gaussian distribution~\cite{Nicolini:2005vd} --- or alternatively by a Lorentzian distribution~\cite{Nozari:2008rc}.
In this way the mass density takes the form $\rho_{\theta}(r)=\frac{M}{(4\pi\theta)^{3/2}}\exp(-r^2/4\theta)$ or $\rho_{\theta}(r)=\frac{M\sqrt{\theta}}{\pi^{3/2}(r^2+\pi\theta)^2}$,
where $\theta$ is the noncommutative parameter and $M$ is the total mass diffused throughout the region of linear size $\sqrt{\theta}$. }
In the present work we aim to analyze the scattering process of a noncommutative Schwarzschild black hole. 
For this purpose we will apply the partial wave approach in order to determine the scattering and absorption cross section at the low frequency limit.}

Several works have been proposed to investigate different aspects of scalar wave scattering by black holes~\cite{Futterman1988}. 
A number of studies have found that  at the low frequency limit  
($ GM\omega\ll 1 $)~\cite{Matzner1977,Westervelt1971,Peters1976,Sanchez1976,Logi1977,Doram2002,Dolan:2007ut,Crispino:2009ki}, 
the differential scattering cross section at small angle limit displays the following result: $d\sigma/d\Omega\approx 16G^2M^2/\vartheta^4 $. 
Besides that, many previous studies~\cite{Churilov1974,Gibbons1975,Page1976} have reported that at the low frequency limit the absorption cross section is equal to the area  of the black hole event horizon $\sigma=4\pi r^2_h=16\pi G^2 M^2$~\cite{Churilov1973}.
In the last few years, an extension of the partial wave method has also been applied in the physics of analogous black holes in fluids~\cite{Crispino:2007zz,Dolan:2009zza,Oliveira:2010zzb,Dolan,ABP2012-1,Anacleto:2015mta,Anacleto:2018acl} as well as in the investigation of scattering due to a non-commutative BTZ black hole~\cite{Brito2015}.
Some studies on the scattering process of massive fields of black holes were also investigated~\cite{Jung2004,Doran2005,Dolanprd2006,Castineiras2007,Benone:2014qaa}. 
In string theory the scalar wave scattering by spherically symmetric $d$-dimensional black holes has also been analyzed~\cite{Moura:2011rr}.
In~\cite{Paik:2017wcy,Kumar:2017hgs} the authors considering a noncommutative Schwarzschild black hole have explored the effect of the noncommutativity on the matter accretion rate.

{As supported by the several results in the literature, e.g. \cite{Rizzo:2006zb,Mustafa:2019eet,rani},  the essential aspects of the noncommutativity approach are not sensitive to using Gaussian or Lorentzian 
distributions, then we shall consider the latter for two main reasons. First, because it is more tractable in some specific analytical steps of the present study and second, because in our formalism the Gaussian terms are highly suppressed --- see below. Thus, in the present paper we are interested in investigating the effect of noncommutativity by considering a Lorentzian smeared mass distribution} on the scattering problem of a massless scalar wave by a noncommutative Schwarzschild black hole.
Thus, by adopting the technique developed in~\cite{Dolan,ABP2012-1,Anacleto:2015mta,Brito2015,Marinho:2016ixt,Anacleto:2017kmg}, we shall focus on the computation of the scattering and absorption cross section for a monochromatic planar wave of neutral massless scalar field impinging upon a noncommutative black hole. 
In our analyzes, we will consider only the long-wavelength regime, in which $ GM\omega\ll 1 $. 
In~\cite{Anacleto:2017kmg} the authors introduced an approximation formula  to compute the phase shift $ \delta_{l}\approx (l-\ell) $ analytically, 
where $ \ell $ is defined by considering only the contributions of $ l $ and $ \omega $ appearing 
in the $ 1/r^ 2 $ term modified after the power series expansion of $ 1/r $.
Therefore, by adopting the approximation formula mentioned above, we find the phase shift analytically in order to compute 
the scattering/absorption cross section for the noncommutative Schwarzschild black hole in the low energy regime.
As a result, we show that in the low-frequency limit the scattering/absorption cross section has its value decreased when we increase the value of the non-commutativity parameter. In addition, we have confirmed this result by solving the problem numerically for arbitrary frequencies.
Moreover also argued that the obtained results are similar to the scattering process of a massless scalar wave by a Reissner-Nordstr\"{o}m black hole~\cite{Crispino:2009ki}, with the difference of the fact that the generated ``effective charge'' in this case is due to space noncommutativity. 
In~\cite{Kim:2008vi} the thermodynamic similarity between the noncommutative Schwarzschild black hole and the Reissner-Nordstr\"{o}m black hole has been investigated.
Other types of black holes that have similarities to the Reissner-Nordstr\"{o}m black hole have been investigated in~\cite{Macedo:2016yyo} and~\cite{deOliveira:2018kcq} by calculating the total absorption cross section and the differential scattering cross section.

The paper is organized as follows. In Sec.~\ref{NSBH} we derive the phase shift and calculate the scattering/absorption cross section for a noncommutative Schwarzschild black hole by considering analytical and numerical analysis. In Sec.~\ref{conclu} we make our final considerations. We shall adopt the natural units $ \hbar=c=k_B=1$ throughout the paper.

\section{Noncommutative Schwarzschild Black Hole}
\label{NSBH}
In this section we aim to derive the phase shift in order to compute the differential scattering/absorption cross section for a noncommutative Schwarzschild black hole by the partial wave method at the low frequency limit. 

\subsection{Scattering/Absorption Cross Section}
The line element of a noncommutative Schwarzschild black hole {due to the Gaussian distribution} is given by~\cite{Nicolini:2005vd}
\begin{eqnarray}
\label{metrnc}
ds^2=f(r)dt^2-\frac{dr^2}{f(r)}-r^2d\Omega^2,
\end{eqnarray}
where
\begin{eqnarray}
\label{metmonopole}
f(r)=1-\frac{4M}{\sqrt{\pi}r}\gamma\left(\frac{3}{2},\frac{r^2}{4\theta}  \right),
\end{eqnarray}
being
\begin{eqnarray}
\label{gamma}
\gamma\left(\frac{3}{2},\frac{r^2}{4\theta}\right)\equiv
\int^{r^2/4\theta}_{0} dt\sqrt{t}e^{-t}=\Gamma\left(\frac{3}{2} \right)- \Gamma\left(\frac{3}{2}, \frac{r^2}{4\theta}\right),
\end{eqnarray}
the lower incomplete Gamma function and $ \Gamma $ the upper incomplete Gamma function.
By imposing the condition $ f(r)=0$, we obtain the event horizon radius given by
\begin{eqnarray}
\label{rtheta}
r_{H}=\frac{2r_s}{\sqrt{\pi}}\gamma\left(\frac{3}{2},\frac{r^2_H}{4\theta}\right),
\end{eqnarray}
{where $r_s=2M$ is the event horizon of the Schwarzschild black hole.}
The above equation can be written in terms of the $ \Gamma $ function as in the following
\begin{eqnarray}
\label{rtheta2}
r_{H}={r_s}\left[1-\frac{2}{\sqrt{\pi}}\Gamma\left(\frac{3}{2},\frac{M^2}{4\theta}\right)\right].
\end{eqnarray}
For $ M^2/\theta \gg 1 $, we have
\begin{eqnarray}
\label{rtheta3}
r_{H}={r_s}\left[1-\frac{M}{\sqrt{\pi\theta}}e^{-M^2/\theta}\right].
\end{eqnarray}
The Hawking temperature for the noncommutative Schwarzschild black hole is given by
\begin{eqnarray}
\label{temp}
T_H=\frac{f^{\prime}(r_H)}{4\pi}=\frac{1}{4\pi r_H}\left(1-\frac{r_H^3}{4\theta^{3/2}}\frac{e^{-r_H^2/\theta}}{\gamma\left({3}/{2},{r^2_H}/{4\theta}\right)}\right).
\end{eqnarray}
We can write the Hawking temperature in the $ {r^2_H}/{4\theta}\gg 1 $ regime, replacing the horizon radius of equation (\ref{rtheta3}) in (\ref{temp}) and keeping the terms up to leadind order in $ \theta $, so we have
\begin{eqnarray}
T_H=\frac{1}{8\pi M}\left[1-\frac{4M^3}{\sqrt{\pi\theta^3}}e^{-M^2/\theta}\right].
\end{eqnarray}
Note that 
by maintaining only the first term we recover the Hawking temperature of the Schwarzschild black hole, which is approximately the case for aforementioned limit. 

Now considering the limit ${r^2}/{4\theta}\gg 1$ into (\ref{gamma}) we have
\begin{eqnarray}
\gamma\left(\frac{3}{2},\frac{r^2}{4\theta}\right)\approx \frac{\sqrt{\pi}}{2}-\frac{r}{2\sqrt{\theta}}e^{-{r^2}/{4\theta}},
\end{eqnarray}
such that we can write the metric (\ref{metrnc}) in the form
\begin{eqnarray}
\label{metrnc2}
ds^2=A(r)dt^2-\frac{dr^2}{A(r)}-r^2d\Omega^2,
\end{eqnarray}
where
\begin{eqnarray}
\label{fmetnc}
A(r)=1-\frac{2M}{r}+\frac{2M}{\sqrt{\pi}}\frac{ e^{-r^2/4\theta}}{\sqrt{\theta}}
\end{eqnarray}

{Similar computations can be done for the Lorentzian distribution. More specifically, we can get the metric for this case by considering the mass density~\cite{Nozari:2008rc}
\begin{eqnarray}
\label{distlornt}
\rho_{\theta}(r)=\frac{M\sqrt{\theta}}{\pi^{3/2}(r^2+\pi\theta)^2}.
\end{eqnarray}
Now the smeared mass distribution function can be obtained by the following integration
\begin{eqnarray}
{\cal M}_{\theta}&=&\int_{0}^{r}\rho_{\theta}(r)4\pi r^2 dr,
\\
&=&\frac{2M}{\pi}\left[\tan^{-1}\left( \frac{r}{\sqrt{\pi\theta}} \right) -\frac{r\sqrt{\pi\theta}}{\pi\theta+r^2}  \right],
\\
&=& M-\frac{4M\sqrt{\theta}}{\sqrt{\pi} r}+ {\cal O}(\theta^{3/2}).
\end{eqnarray}
where in the last step we consider the limit $r^2\gg\pi\theta$. Consequently, by considering the above modified mass, the metric function of the noncommutative black hole can now be given by
\begin{eqnarray}
\label{metlorntz}
A(r)=1-\frac{2{\cal M}_{\theta}}{r}=1-\frac{2M}{r}+\frac{8M\sqrt{\theta}}{\sqrt{\pi} r^2}+ {\cal O}(\theta^{3/2}).
\end{eqnarray}
}{Now comparing the metric functions (\ref{fmetnc}) and (\ref{metlorntz}) we choose the latter in order to simplify the computation of the phase shift in the calculation of the scattering/absorption cross section. Furthermore, in such a computation, a power expansion $1/r$ for large distance is considered, such that the Gaussian term in the function (\ref{fmetnc}) is highly suppressed in comparison with the Lorentzian distribution and then no sensitive noncommutative effects are expected to appear from it.}

By considering the metric (\ref{metlorntz}) we obtain the external and internal event horizons at the small $\theta$ limit.  By keeping terms up to first order in $ \sqrt{\theta}$, one finds respectively
\begin{eqnarray}
\tilde{r}_{h}&=&2M-4\sqrt{\frac{\theta}{\pi}}+ \cdots=2M\left(1-\frac{2}{M}\sqrt{\frac{\theta}{\pi}} \right)+ \cdots ,
\\
r_{\theta}&=&4\sqrt{\frac{\theta}{\pi}}+ \cdots .
\end{eqnarray}

In the next step we consider the case of the massless scalar field equation to describe the scattered wave. In this case the massless scalar waves are described by the Klein-Gordon equation in the background (\ref{metrnc2}),  given by 
\begin{eqnarray}
\dfrac{1}{\sqrt{-g}}\partial_{\mu}\Big(\sqrt{-g}g^{\mu\nu}\partial_{\nu}\Phi\Big)=0 .
\end{eqnarray}
Now we can make a separation of variables into the equation above as follows
\begin{eqnarray}
\Phi_{\omega l m}({\bf r},t)=\frac{R_{\omega l}(r)}{r}Y_{lm}(\theta,\phi)e^{-i\omega t},
\end{eqnarray}
where $ \omega $ is the frequency and $Y_{lm}(\theta,\phi) $ are the spherical harmonics.

In this case, the equation for $ R_{\omega l}(r) $ can be written as 
\begin{eqnarray}
\label{eqrad}
A(r)\dfrac{d}{dr}\left(A(r)\dfrac{dR_{\omega l}(r)}{dr} \right) +\left[ \omega^2 -V_{eff} \right]R_{\omega l}(r)=0,
\end{eqnarray}
where $ V_{eff} $ is the effective potential, given by
\begin{eqnarray}
V_{eff}=\frac{A(r)}{r}\frac{dA(r)}{dr}+\frac{A(r)l(l+1)}{r^2}.
\end{eqnarray}
Next, we  consider a new radial function, $ \psi(r)=A^{1/2}(r)R(r) $, so we have
\begin{eqnarray}
\label{eqradpsi}
\dfrac{d^2\psi(r)}{dr^2}+U(r) \psi(r) = 0,
\end{eqnarray}
where
\begin{eqnarray}
\label{poteff}
U(r)=\dfrac{[A'(r)]^2}{4 A^2(r)} - \dfrac{A''(r)}{2A(r)} + \dfrac{\omega^2}{A^2(r)} - \dfrac{V_{eff}}{A^2(r)},
\end{eqnarray} 
\begin{eqnarray}
&&A'(r)=\frac{dA(r)}{dr} =\frac{2M}{r^{2}}-\frac{16\sqrt{\theta}M}{\sqrt{\pi}r^3} , 
\quad \quad A''(r)=\frac{d^2A(r)}{dr^2}= - \frac{4M}{r^{3}} +\frac{48\sqrt{\theta}M}{\sqrt{\pi}r^4}.
\end{eqnarray}
Now, we can perform a power series expansion in $1/r$ for the potential $ U(r) $, so the Eq.~(\ref{eqradpsi}) becomes
\begin{eqnarray}
\frac{d^2\psi(r)}{dr^2}+\left[\omega^2+ {\cal U}(r)\right] \psi(r) = 0,
\end{eqnarray}
{ where now keeping corrections up to first order in $\sqrt{\theta}$ we have 
\begin{eqnarray}
\label{pot1}
{\cal U}(r)= \frac{4M\omega^2}{r}+\frac{12\ell^2}{r^2}
+\frac{32M\omega^2}{r^2}\sqrt{\frac{\theta}{\pi}}+\cdots,
\end{eqnarray}
}
 and we define 
\begin{eqnarray}
\label{ell}
\ell^2\equiv-\frac{(l^2+l)}{12}+M^2\omega^2\left(1-\frac{4}{M}\sqrt{\frac{\theta}{\pi}}\right).
\end{eqnarray}
{Here $ \ell^2 $ was defined as the change of the coefficient of $1/r ^ 2$ (containing only the contributions involving $ l , M, \theta$ and $ \omega $) that arises after the realization of the power series expansion in $1/r $ in Eq. (\ref{eqradpsi}).}
Notice that when $ r \rightarrow \infty $ the potential $ {\cal U}(r) \rightarrow 0 $ and the suitable asymptotic behavior is satisfied.

The phase shift $ \delta_{l} $ can be obtained applying the following approximated formula
\begin{eqnarray}
\label{formapprox}
\delta_l\approx 2(l-\ell)=2\left(l - \sqrt{-\frac{(l^2+l)}{12}
+M^2\omega^2\left(1-\frac{4}{M}\sqrt{\frac{\theta}{\pi}}\right)}\right).
\end{eqnarray}
In the limit $ l\rightarrow 0 $ we obtain
\begin{eqnarray}
\label{phase2}
\delta_l=-{2M\omega}\left(1-\frac{4}{M}\sqrt{\frac{\theta}{\pi}}\right)^{1/2}+{\cal O}(l)
\approx-{2M\omega}\left(1-\frac{2}{M}\sqrt{\frac{\theta}{\pi}}\right).
\end{eqnarray}
Note that in the limit $ l\rightarrow 0 $ the phase shifts tend to a non-zero term.
In this way, knowing the expression for the phase shifts we can determine the scattering amplitude. 
Therefore, to determine the differential scattering cross section, we will use the following equation~\cite{Yennie1954,Cotaescu:2014jca}
\begin{eqnarray}
\label{espalh}
\dfrac{d\sigma}{d\vartheta}=\big|f(\vartheta) \big|^2=\Big| \frac{1}{2i{\omega}}\sum_{l=0}^{\infty}(2l+1)\left(e^{2i\delta_l} -1 \right)
\frac{P_{l}\cos\vartheta}{1-\cos\vartheta}\Big|^2.
\end{eqnarray}
In the limit of small angles it becomes
\begin{eqnarray}
\label{espalh2}
\frac{d\sigma}{d\vartheta}&=&\frac{4}{\omega^2\vartheta^4}\Big|\sum_{l=0}^{\infty}(2l+1)\sin(\delta_{l})
{P_{l}\cos\vartheta}\Big|^2
\\
&=&\frac{16M^2}{\vartheta^4}\left( 1-\frac{4}{M}\sqrt{\frac{\theta}{\pi}}\right)\Big|\sum_{l=0}^{\infty}(2l+1)
{P_{l}\cos\vartheta}\Big|^2.
\end{eqnarray}
Hence the differential scattering cross section in this limit and for $ l=0 $ is given by
\begin{eqnarray}
\frac{d\sigma}{d\vartheta}\Big |^{\mathrm{l f}}_{\omega\rightarrow 0}
=\frac{16M^2}{\vartheta^4}\left( 1-\frac{4}{M}\sqrt{\frac{\theta}{\pi}}\right)+\cdots.
\end{eqnarray}
The dominant term is modified by the noncommutative parameter $ \theta $. 
Thus, we verified that the differential cross section is decreased by the effect of noncommutativity.
As $ \theta=0 $ we obtain the result for the Schwarzschild black hole case.

{Now we will determine the absorption cross section for a noncommutative Schwarzschild black hole at the low frequency  limit.}
As is well known in quantum mechanics, the total absorption cross section can be computed by means of the following relation
\begin{eqnarray}
\sigma_{abs}
=\frac{\pi}{\omega^2}\sum_{l=0}^{\infty}(2l+1)\Big(\big|1-e^{2i\delta_l}\big|^2\Big)
=\frac{4\pi}{\omega^2}\sum_{l=0}^{\infty}(2l+1)\sin^2(\delta_{l}).
\end{eqnarray}
For the phase shift $ \delta_l $ of the Eq. (\ref{phase2}), we obtain in the limit $ \omega\rightarrow 0  $ $(l=0):$
\begin{eqnarray}
\label{abs1}
\sigma_{abs}^{\mathrm{l f}}
&=& 16\pi{M^2}\left( 1-\frac{4}{M}\sqrt{\frac{\theta}{\pi}}\right)={\cal A}_{Sch}\left( 1-\frac{4}{M}\sqrt{\frac{\theta}{\pi}}\right),
\end{eqnarray}
where $ {\cal A}_{Sch} =4\pi r^2_s=16\pi{M^2}$ is the area of the event horizon of the Schwarzschild black hole.
Notice that the absorption is decreased due to the contribution of the noncommutativity. {Equation (\ref{abs1}) can also be rewritten in terms of the event horizon area of the non-commutative black hole as follows
\begin{eqnarray}
\label{abs1-2}
\sigma_{abs}^{\mathrm{l f}}
\approx4\pi\left[2M\left( 1-\frac{2}{M}\sqrt{\frac{\theta}{\pi}}\right)\right]^2=4\pi{\tilde{r}_{h}^2}={\cal A}_{ncSch},
\end{eqnarray}
where $ {\cal A}_{ncSch} $ is the area of the event horizon of the noncommutative Schwarzschild black hole.
Here we have kept terms up to first order in $ \sqrt{\theta} $.}
Furthermore, our results for absorption show concordance with the universality property of the absorption cross section which is always proportional to the area of the event horizon at low-frequency limit~\cite{Das:1996we}.
{In addition, in Fig. \ref{l0} we show the graph for the mode $ l=0 $ of the absorption cross section that was obtained by numerically solving the radial equation (\ref{eqrad}) for arbitrary frequencies. 
By analyzing the behavior of Fig. \ref{l0} we can observe that the non-commutative parameter $ \theta $ introduces a charge effect when compared with the graph of the absorption of the Reissner-Nordstr\"{o}m black hole~\cite{Crispino:2009ki}.

The line element for the Reissner-Nordstr\"{o}m black hole is given by equation (\ref{metrnc2}) with
\begin{eqnarray}
\label{metrn}
A_{RN}(r)= 1-\frac{2M}{r}+\frac{Q^2}{r^2}.
\end{eqnarray}
Here we should emphasize that the noncommutative parameter is related to the charge of the Reissner-Nordstr\"{o}m black hole by the relation
\begin{eqnarray}
\label{charge}
\sqrt{\theta}=\frac{\sqrt{\pi}Q^2}{8M}.
\end{eqnarray}
From relation (\ref{charge}) note that the square root of the noncommutative parameter is proportional to the square of the electric charge of the Reissner-Nordstr\"{o}m black hole.
\begin{figure}[htbh]
 \centering
 \subfigure[]
 {\includegraphics[scale=0.28]{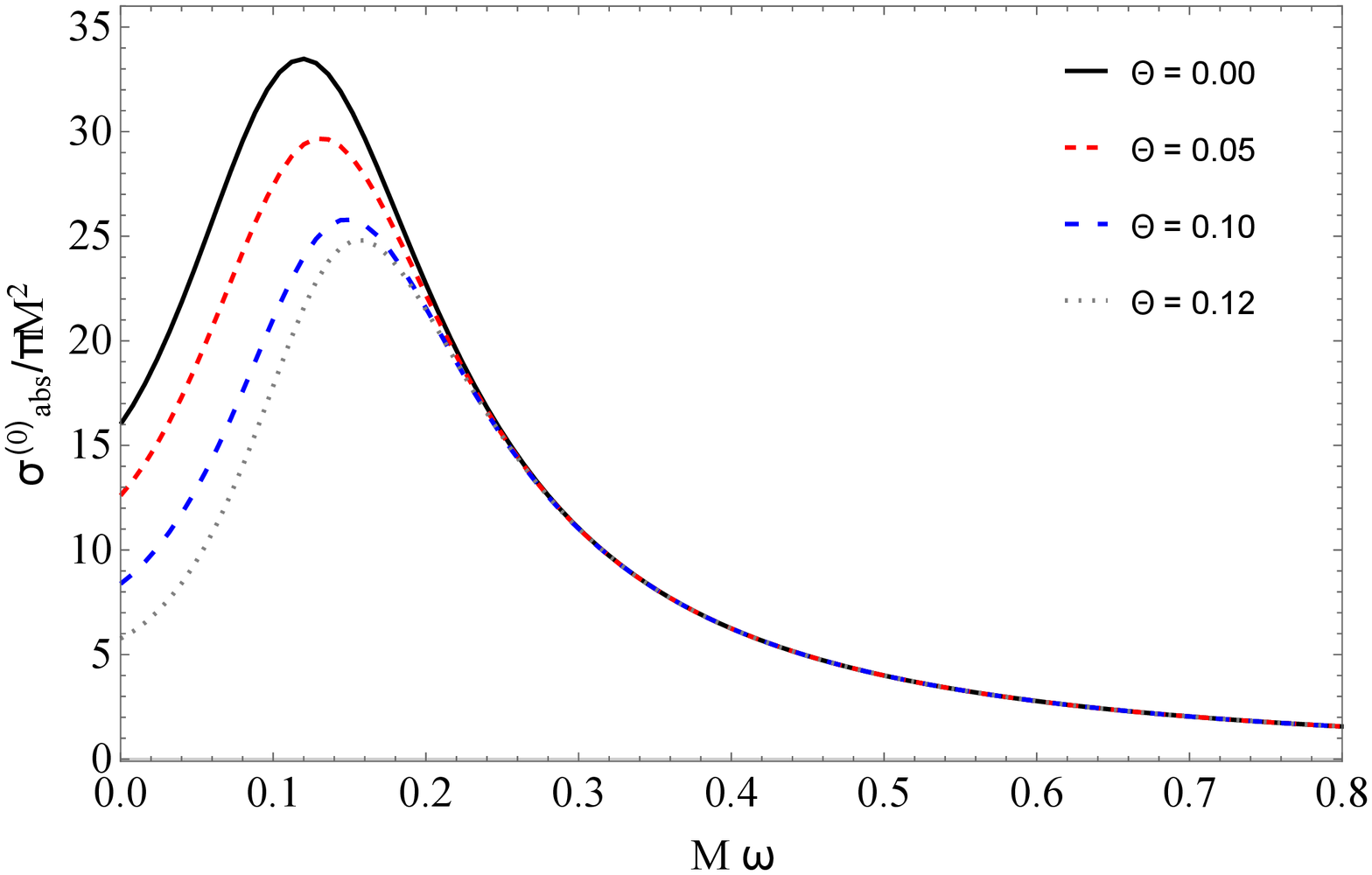}\label{l0a}}
 \qquad
 \subfigure[]
 {\includegraphics[scale=0.28]{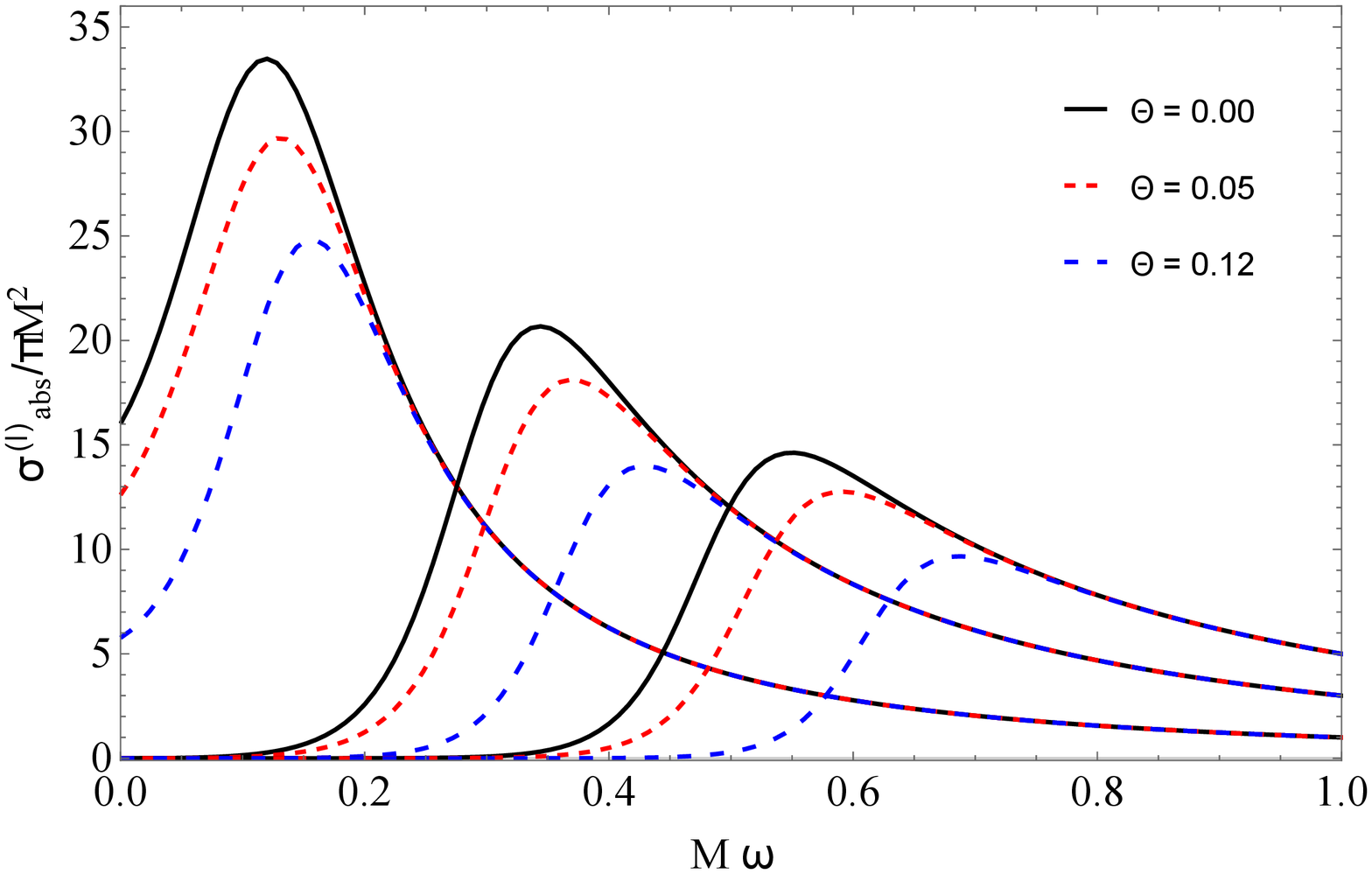}\label{l0b}}
 \caption{Partial absorption cross section to the modes (a) $l=0$ and (b) $ l=0,1,2 $. }
\label{l0}
\end{figure}

\begin{figure}[htbh]
 \centering
 \subfigure[]
 {\includegraphics[scale=0.4]{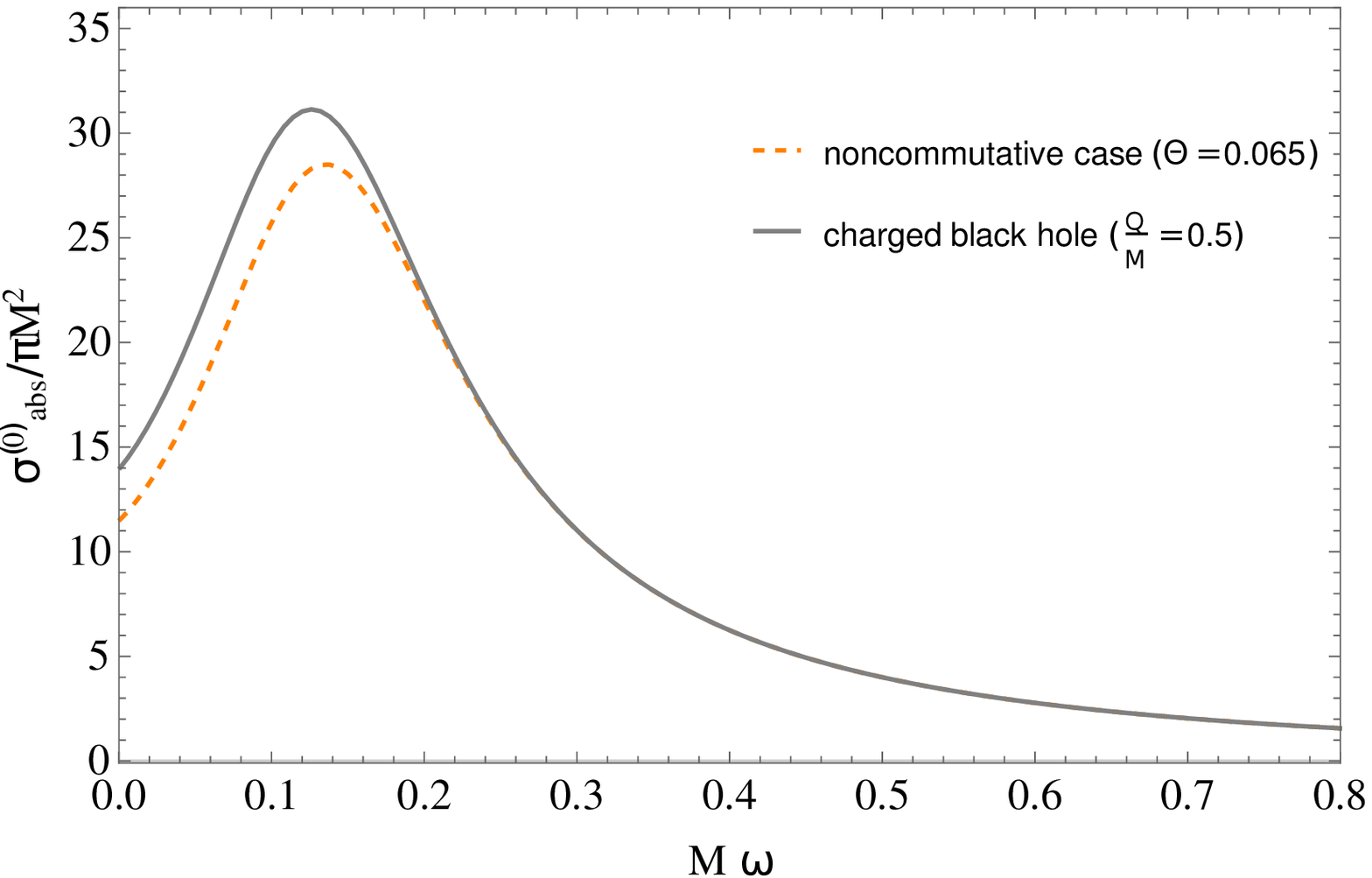}\label{l0nca}}
 \qquad
 \subfigure[]
 {\includegraphics[scale=0.4]{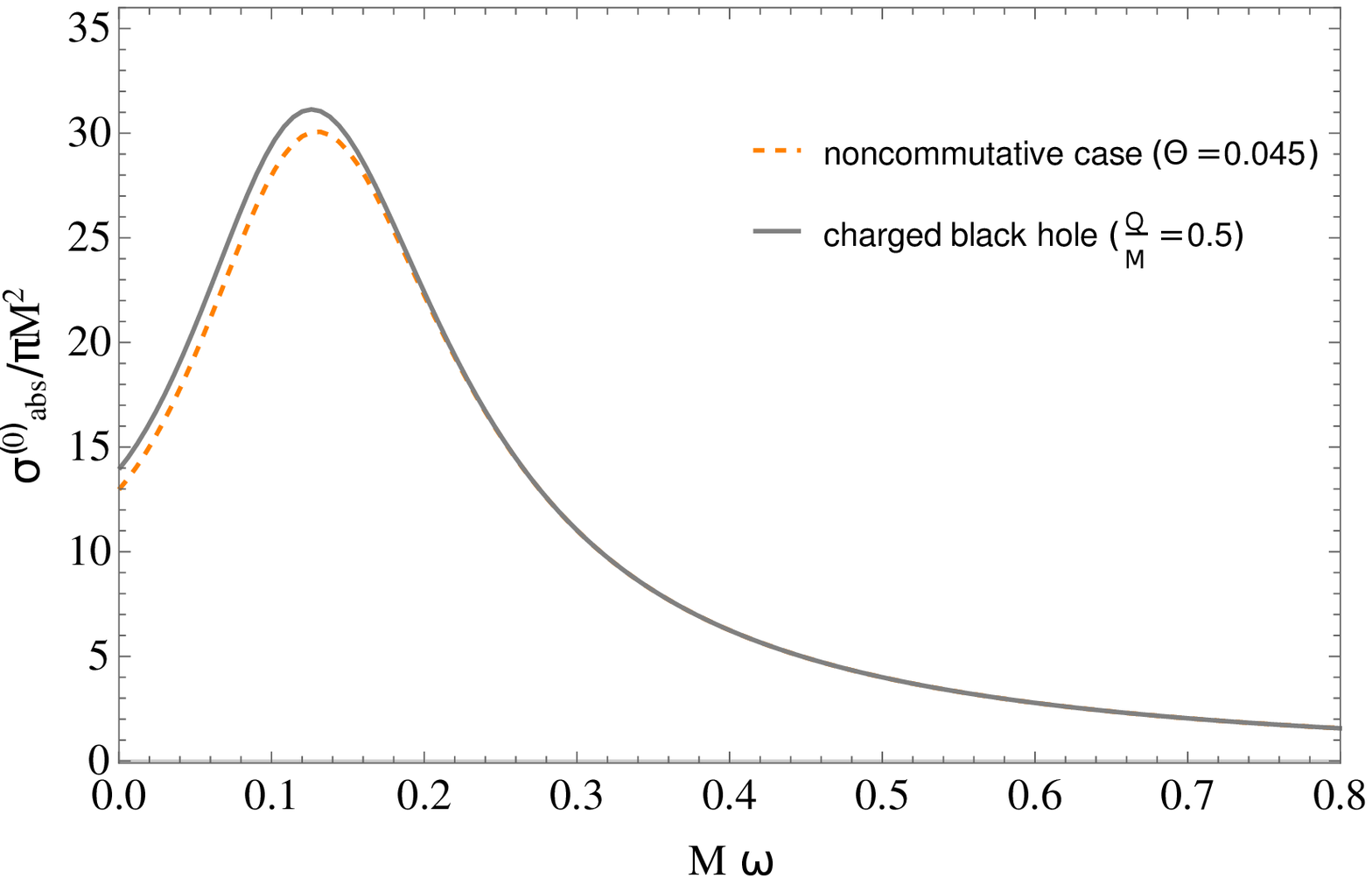}\label{l0ncb}}
 \caption{Partial absorption cross section for mode $ l=0 $ of the Reissner-Nordstr\"{o}m black hole with $ Q/M=0.5 $ compared with noncommutative black hole with $ \Theta  = 0.065$ (a) and $\Theta = 0.045$ (b). }
\label{l0nc}
\end{figure}

\begin{figure}[htbh]
 \centering
 \subfigure[]
 {\includegraphics[scale=0.4]{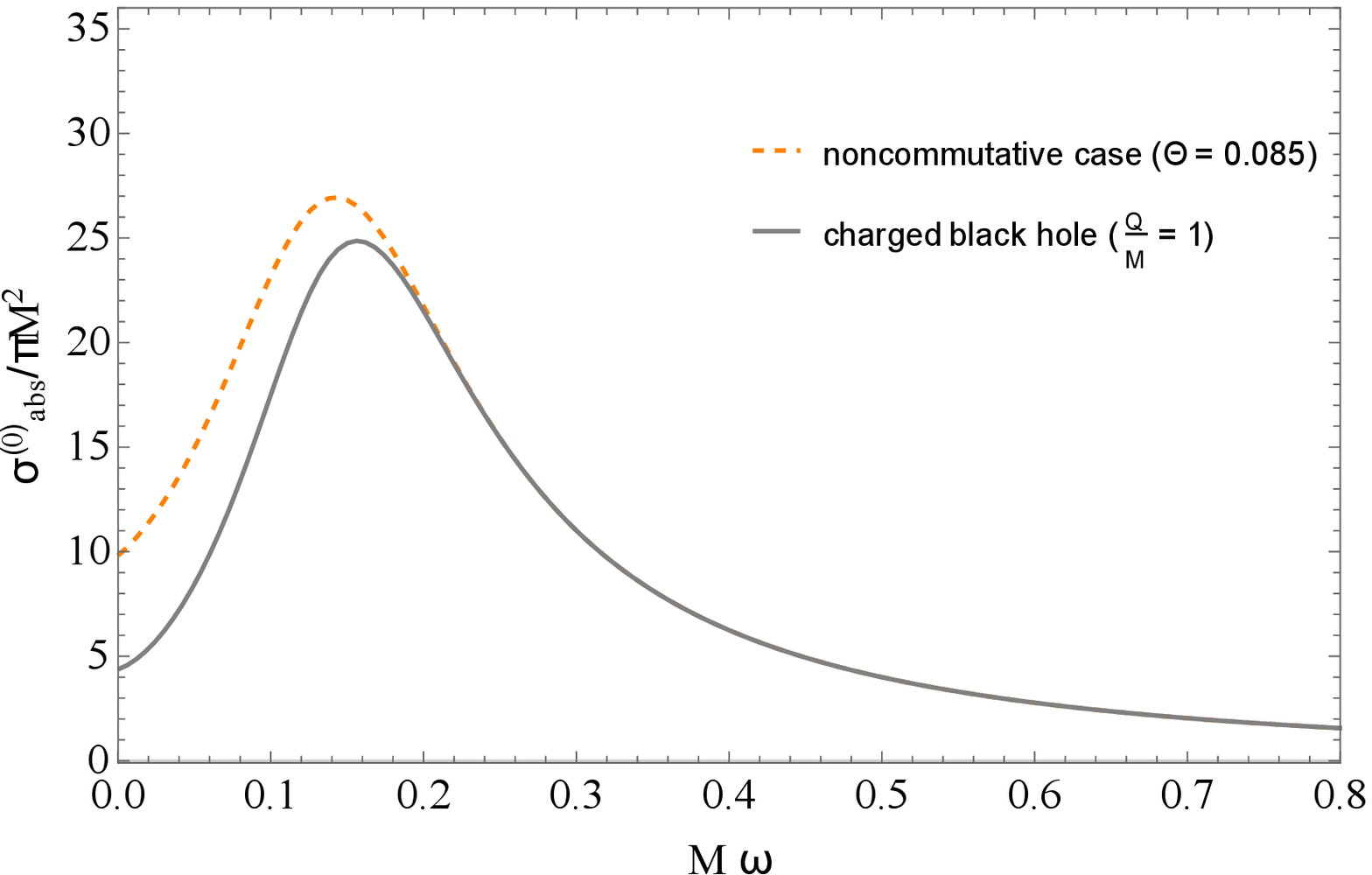}\label{l0exta}}
 \qquad
\subfigure[]
 {\includegraphics[scale=0.4]{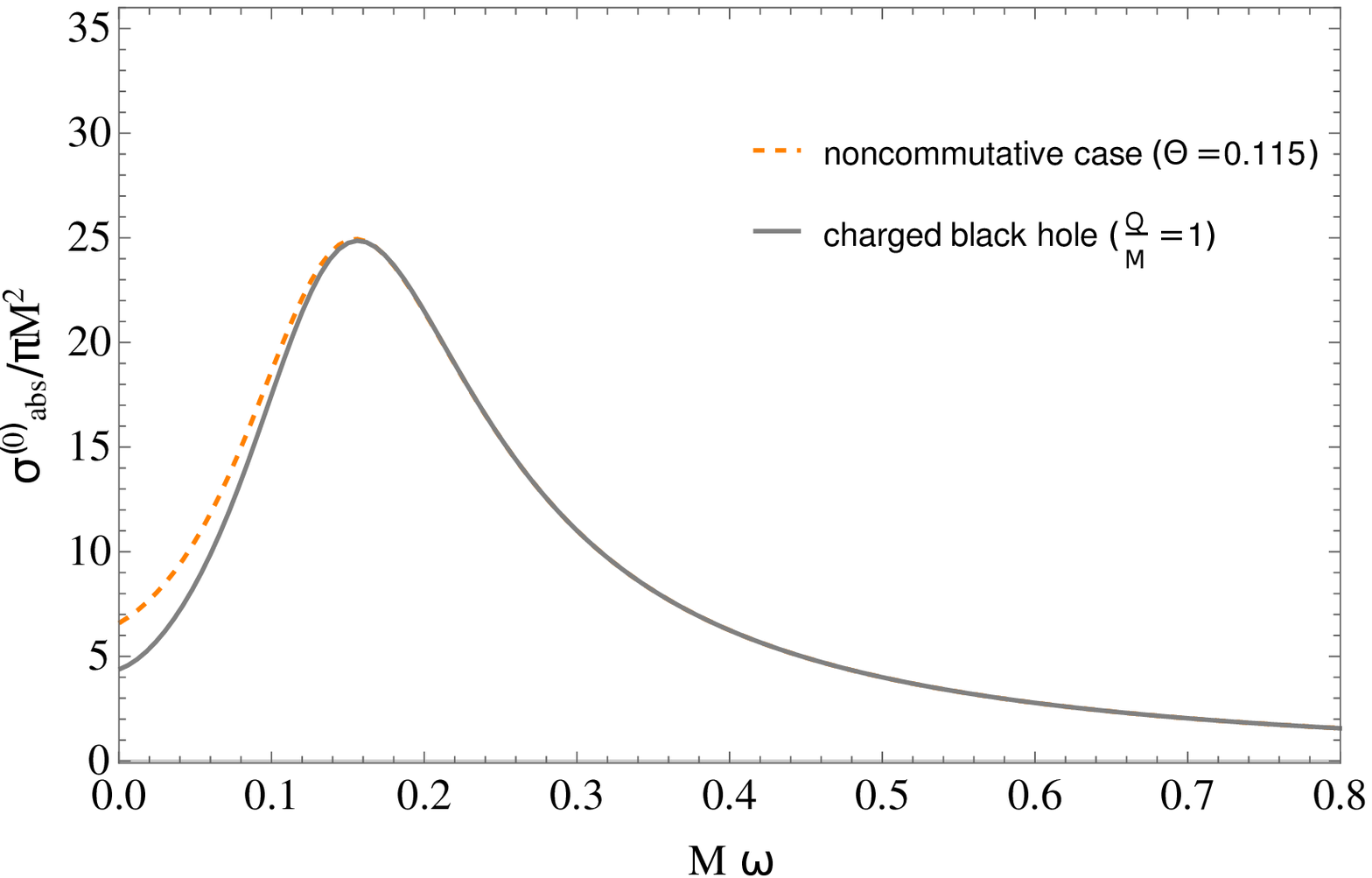}\label{l0extb}}
 \caption{Partial absorption cross section for mode $ l=0 $ of the extremal Reissner-Nordstr\"{o}m black hole compared with noncommutative black hole with $ \Theta  = 0.085$ (a) and $\Theta = 0.115$ (b). }
\label{l0extreme}
\end{figure}

\begin{figure}[htbh]
 \centering
 \subfigure[]
 {\includegraphics[scale=0.28]{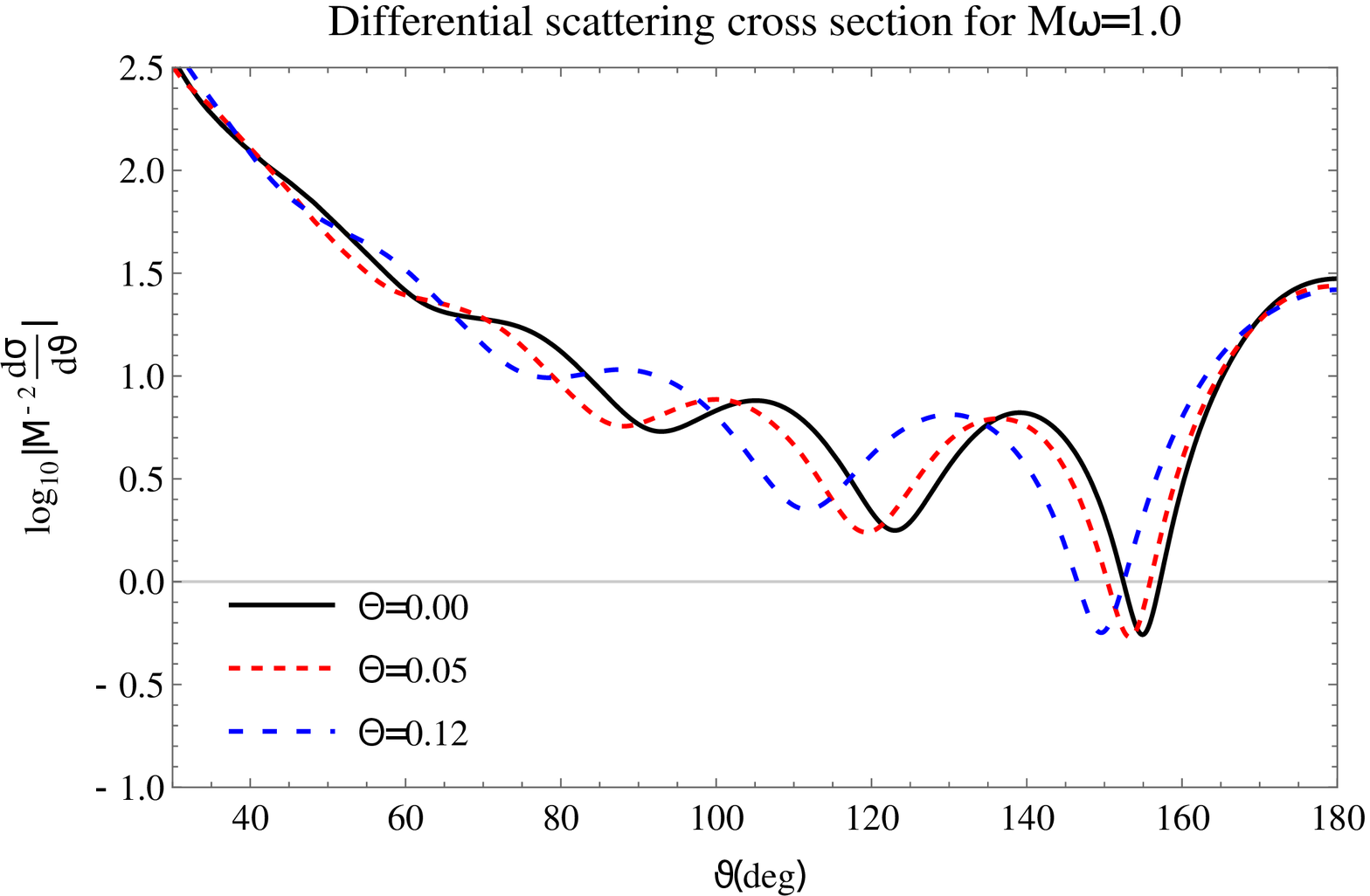}\label{sct1}}
 \qquad
 \subfigure[]
 {\includegraphics[scale=0.28]{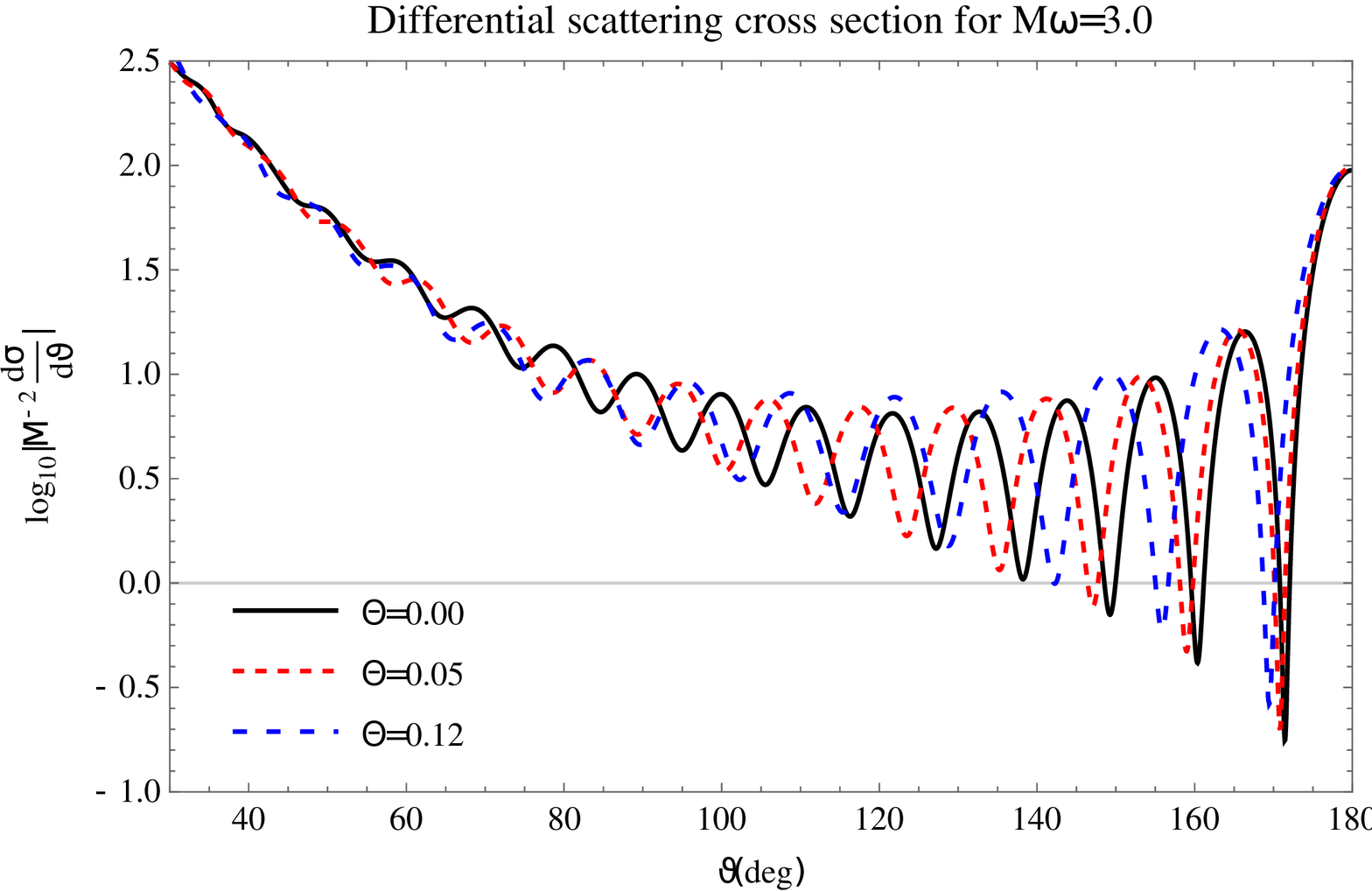}\label{sct2}}
 \caption{Differential scattering cross section for (a) $ M\omega=1.0 $ and (b) $ M\omega=3.0$.}
\label{sct}
\end{figure}

\begin{figure}[htbh]
 \centering
{\includegraphics[scale=0.3]{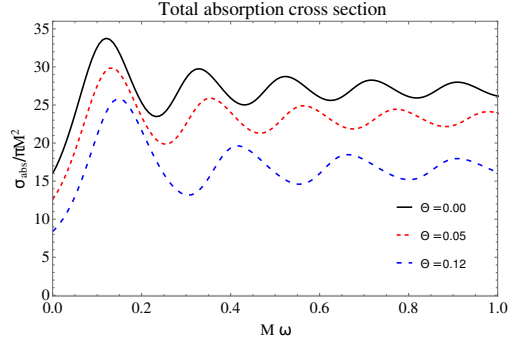}}
 \caption{Comparison between the total absorption cross section
of the noncommutative black hole and Schwarzschild black hole. }
\label{total}
\end{figure}

\begin{figure}[htbh]
 \centering
 \subfigure[]
 {\includegraphics[scale=0.4]{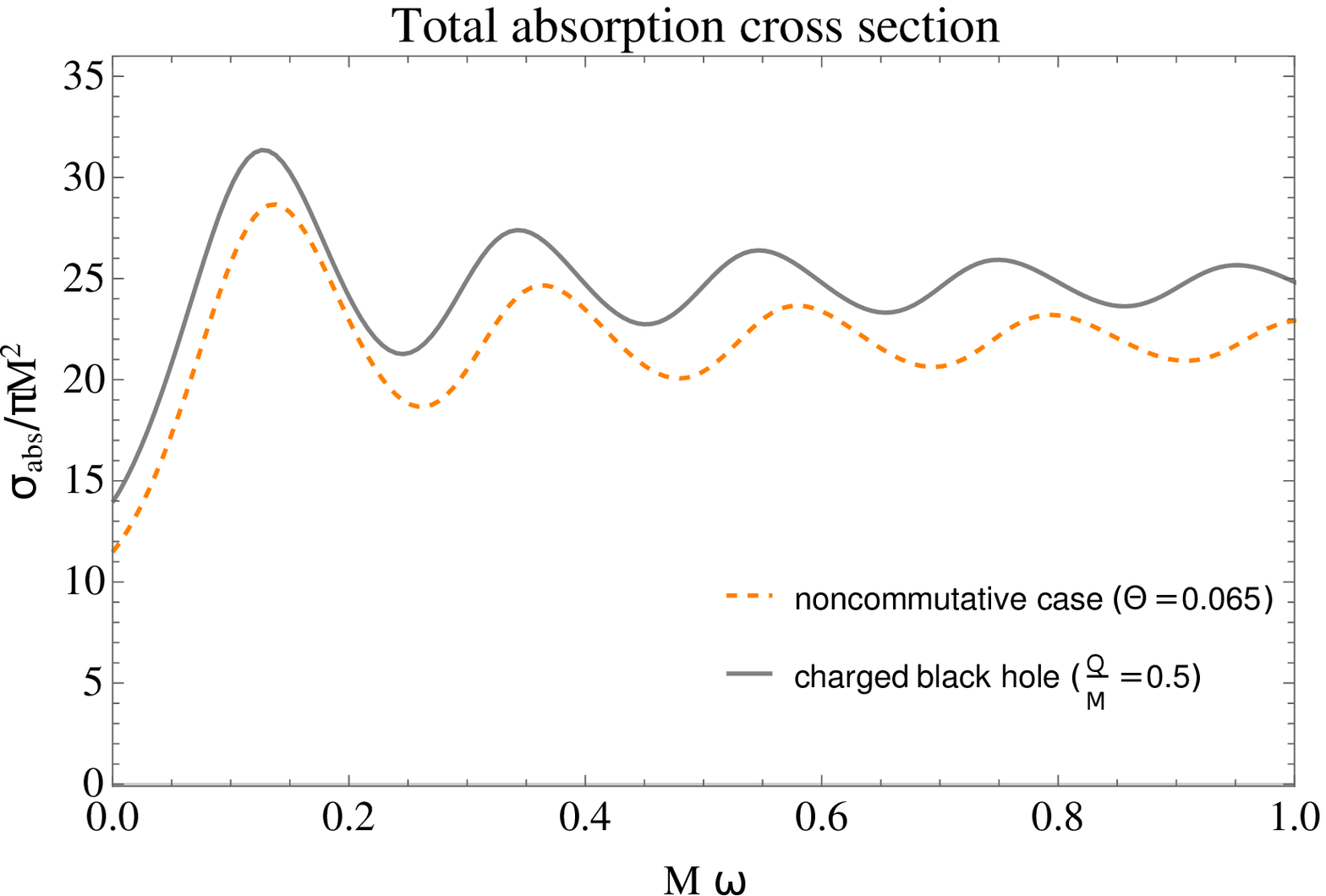}\label{tnca}}
 \qquad
 \subfigure[]
 {\includegraphics[scale=0.4]{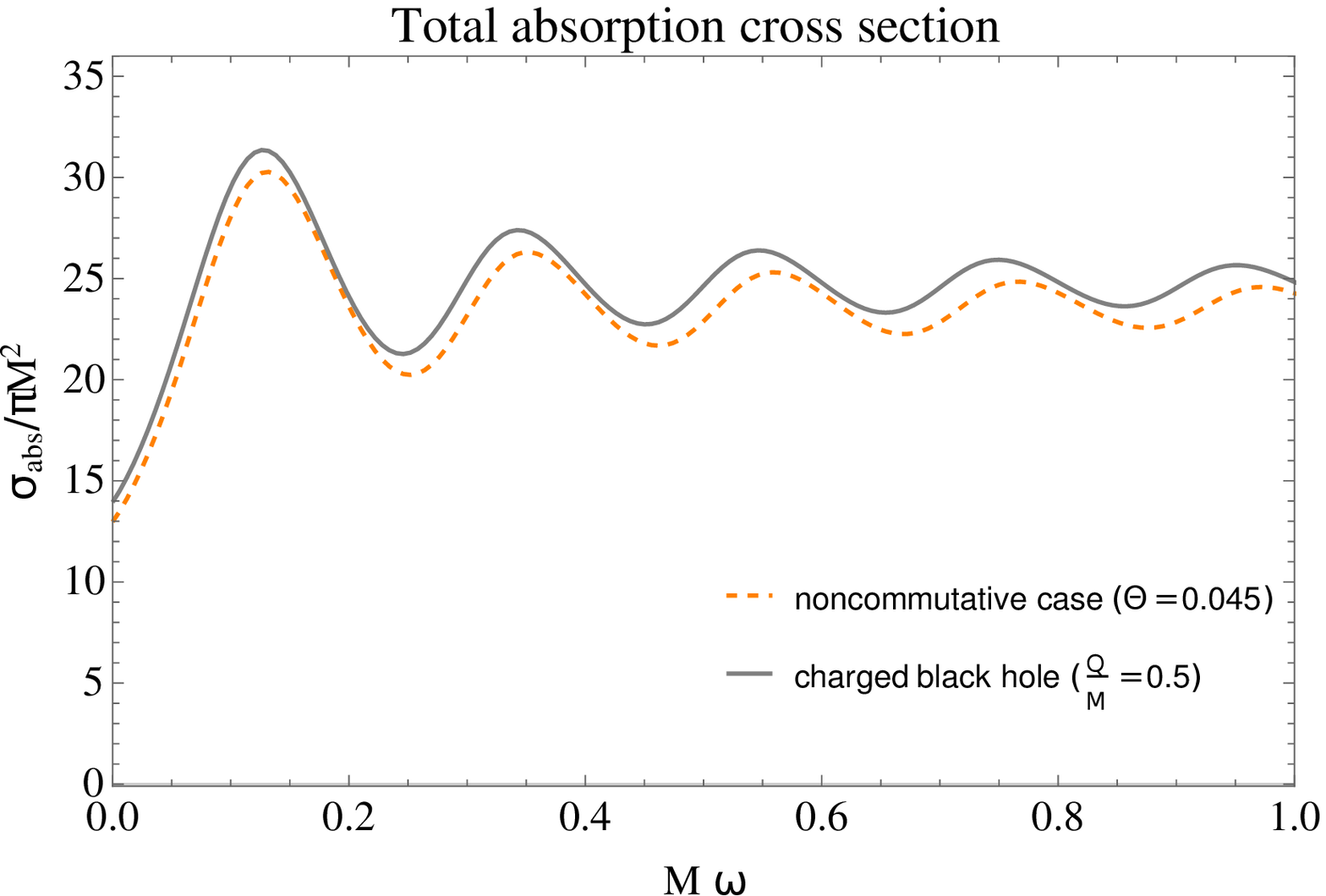}\label{tncb}}
 \caption{ Total absorption cross section of the Reissner-Nordstr\"{o}m black hole with $ Q/M=0.5 $ compared with noncommutative black hole with $ \Theta  = 0.065$ (a) and $\Theta = 0.045$ (b). }
\label{totalnc}
\end{figure}

\begin{figure}[htbh]
 \centering
 \subfigure[]
 {\includegraphics[scale=0.40]{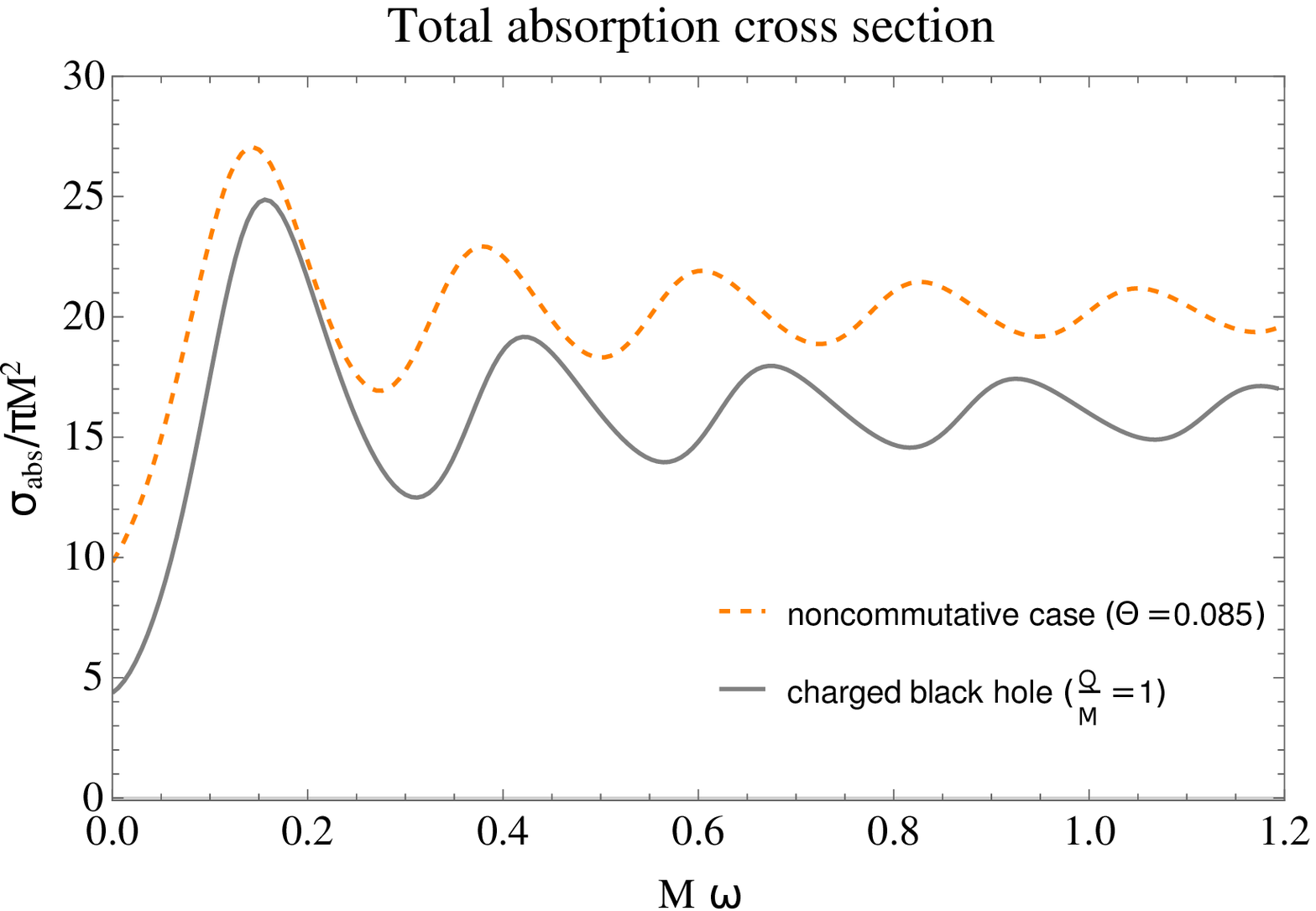}\label{texta}}
 \qquad
 \subfigure[]
 {\includegraphics[scale=0.40]{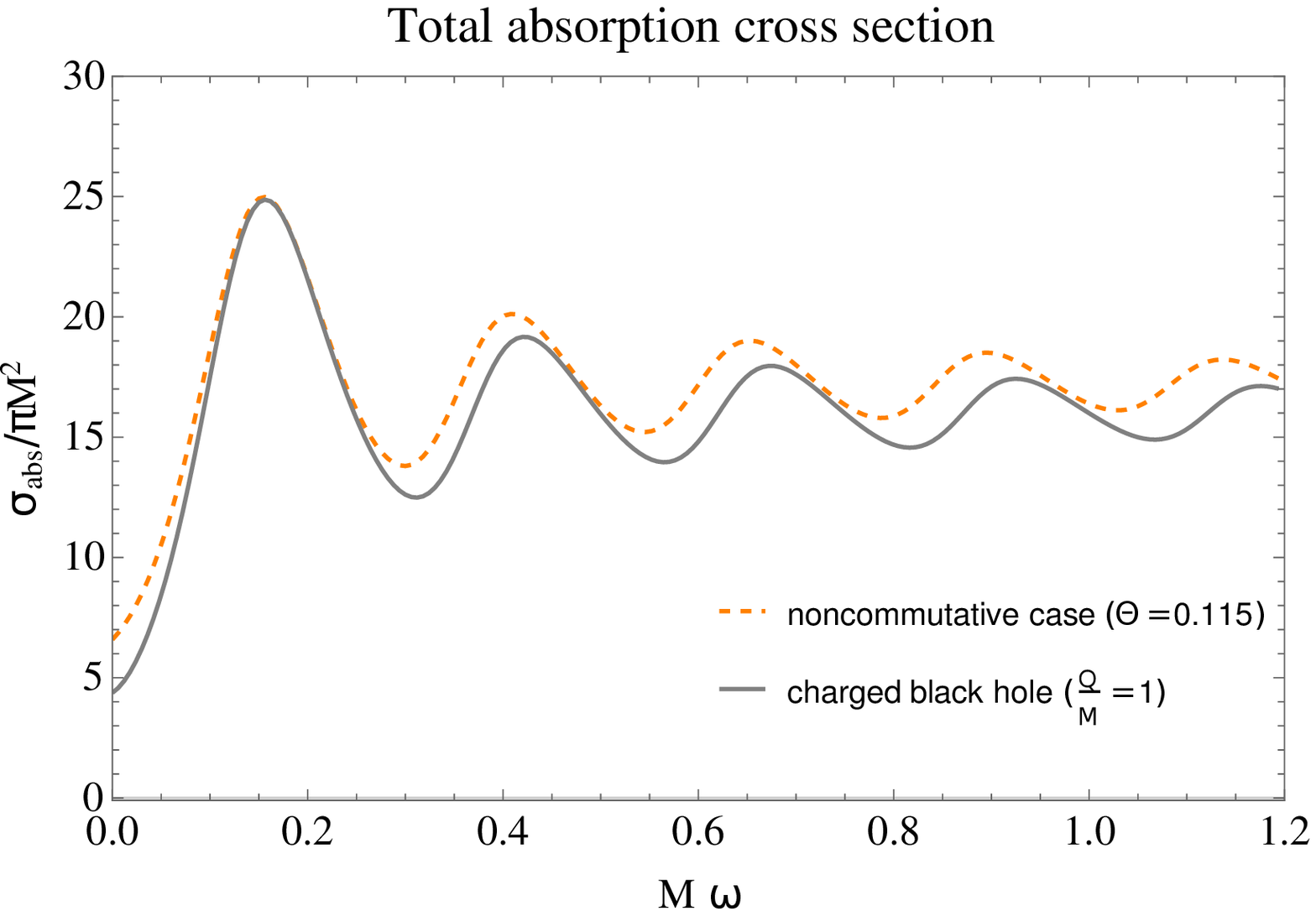}\label{textb}}
 \caption{ Total absorption cross section of the extremal  Reissner-Nordstr\"{o}m black hole compared with noncommutative black hole with
$ \Theta  = 0.85$ (a) and $\Theta = 0.115$ (b). }
\label{totalext}
\end{figure}

\subsection{Numerical Results}
 
Now we present the numerical results of the partial absorption cross section as a function of arbitrary frequencies obtained through the 
numerical procedure as described in~\cite{Dolan:2012yc} by considering arbitrary values of $ 0\leq \Theta\leq 0.125 $, where $\Theta=\sqrt{\theta}/(M\sqrt{\pi})$. The graphs for $ \Theta=0 $ (Schwarzschild case), $ \Theta=0.05 $ and $ \Theta=0.065 $ (noncommutative case) 
 and $\Theta=0.125$ (extreme case) are shown in the figures (Fig.\ref{l0}--Fig.\ref{totalext}). 

In Fig.~\ref{l0a}, we plot the partial absorption cross section for the $ l=0 $ mode. 
We can see by comparing the curves for different values of $\Theta$ that the absorption is decreased due to the contribution of the noncommutativity. 
Moreover, when $ M\omega\rightarrow 0 $ the absorption tends to a nonzero value and when $ M\omega $ increases it tends to zero.
For $ \Theta =0$ the graph shows the result of the partial absorption for the Schwarzschild black hole. 
Thus for non-zero values of $ \Theta $, the partial absorption for the noncommutative Schwarzschild black hole is decreased in relation to the Schwarzschild black hole. 

In Fig.~\ref{l0b}, we plot the contribution of the partial absorption  to the modes $ l=0,1,2 $. We observe that as we increase the values of $\Theta $ the amplitude decreases.

We compared the cases of the non-commutative black hole ($ \Theta=0.045 $ and $\Theta=0.065$) and the Reissner-Nordstr\"{o}m black hole with $ Q/M=0.5$ that is presented in Fig.~\ref{l0nc}. 
We can observe that considering the values of $ \Theta=0.065 $ and $ Q/M=0.5 $ which corresponds to half of the interval in the respective cases the amplitudes do not match. However, the graphs coincide for the values $ Q/M=0.5 $ and $ \Theta=0.031 $ 
as shown in the graphs in Fig.~\ref{l0ncb}.

In Fig.~\ref{l0extreme} we compared the partial absorption cross section for mode $ l=0 $ of the extremal Reissner-Nordstr\"{o}m black hole with 
noncommutative black hole (for $ \Theta=0.085 $ and $\Theta=0.115 $).

In Fig.~\ref{sct} we plot the differential scattering cross section of a noncommutative black hole for the massless scalar field for (a) $ M\omega =1.0 $ and (b) $ M\omega=3.0 $.
In analyzing Fig.~\ref{sct}, we can observe that the effect of noncommutativity on the differential scattering cross section is more significant at large angles.

A comparison of the total absorption cross section for the non-commutative black hole ($\Theta=0.05$ and $ \Theta=0.12 $) and the Schwarzschild black hole ($\Theta=0.0 $) is shown in Fig.~\ref{total}. We observe that when we increase the value of parameter $ \Theta $ the total absorption cross section is reduced as previously noted in the case of partial absorption in the graph of Fig.~\ref{l0}.

Figure~\ref{totalnc} shows the comparison of the total absorption cross section for the noncommutative black hole ($\Theta=0.045$ and $ \Theta=0.065 $) and the Reissner-Nordstr\"{o}m black hole ($ Q/M=0.5$).

Finally, in Fig.~\ref{totalext} we plot the total absorption cross section and  we compared the cases of the non-commutative black hole 
($\Theta=0.085$ and $ \Theta=0.115 $) with the extremal Reissner-Nordstr\"{o}m black hole ($ Q/M=1.0$).

Other types of black holes that have similarities to the Reissner-Nordstr\"{o}m black hole have been analyzed in~\cite{Macedo:2016yyo} and~\cite{deOliveira:2018kcq}.

\section{Conclusions}
\label{conclu}
In summary,  in this work we have explored the scattering of scalar waves by a noncommutative Schwarzschild black hole at the low frequency limit by using analytical methods  and at arbitrary frequencies by numerical methods.
We have determined the phase shift analytically and have used this phase shift to obtain the scattering and absorption cross section at the low frequency limit.
We have obtained that at the small angle limit the dominant term  of the differential scattering cross section is modified by the presence of the $ \theta $ parameter of the noncommutativity.  
We showed that the result for the differential scattering cross section as well as the absorption cross section is decreased due to the noncommutativity effect. 
In addition, we have solved numerically the radial equation in order to calculate the partial absorption cross section for arbitrary frequencies.
As a result we have shown that the absorption has its value decreased as we increase the value of the noncommutative parameter $\Theta$.
Therefore we have verified that the parameter $\Theta$ simulates the effect of an ``effective charge" in the model.
Therefore, the results obtained are similar to the scattering process of a massless scalar wave by a Reissner-Nordstr\"{o}m black hole, with the difference being due to the fact that the generated charge in this case is a consequence of  the space noncommutativity. 
We also have identified that the square root of the noncommutative parameter is proportional to the square of the electric charge.

\acknowledgments

We would like to thank CNPq, CAPES and PRONEX/CNPq/FAPESQ-PB (Grant no. 165/2018), for partial financial support. MAA, FAB and EP acknowledge support from CNPq (Grant nos. 306962/2018-7, 312104/2018-9, 304852/2017-1).


\begin{thebibliography}{100}

\bibitem{Nicolini:2008aj} 
  P.~Nicolini,
  Int.\ J.\ Mod.\ Phys.\ A {\bf 24}, 1229 (2009)
  doi:10.1142/S0217751X09043353
  [arXiv:0807.1939 [hep-th]].
  
\bibitem{Szabo:2006wx} 
  R.~J.~Szabo,
  Class.\ Quant.\ Grav.\  {\bf 23}, R199 (2006)
  doi:10.1088/0264-9381/23/22/R01
  [hep-th/0606233].

\bibitem{Smailagic:2003yb} 
  A.~Smailagic and E.~Spallucci,
  J.\ Phys.\ A {\bf 36}, L467 (2003)
  doi:10.1088/0305-4470/36/33/101
  [hep-th/0307217].
  
\bibitem{Smailagic:2003rp} 
  A.~Smailagic and E.~Spallucci,
  J.\ Phys.\ A {\bf 36}, L517 (2003)
  doi:10.1088/0305-4470/36/39/103
  [hep-th/0308193].  

\bibitem{Nicolini:2005vd} 
  P.~Nicolini, A.~Smailagic and E.~Spallucci,
  Phys.\ Lett.\ B {\bf 632}, 547 (2006)
  doi:10.1016/j.physletb.2005.11.004
  [gr-qc/0510112].

\bibitem{Nozari:2008rc} 
  K.~Nozari and S.~H.~Mehdipour,
  Class.\ Quant.\ Grav.\  {\bf 25}, 175015 (2008)
  doi:10.1088/0264-9381/25/17/175015
  [arXiv:0801.4074 [gr-qc]]. 


\bibitem{Frolov} A.V. Frolov, K.R. Kristjansson, L. Thorlacius et al, Phys. Rev. D {\bf 72}, 021501 (2005), 
[hep-th/0504073]; 

\bibitem{Townsend1997} P. K. Townsend, Black holes: Lecture notes, (University of Cambridge, Cambridge, 1997) [gr-qc/9707012]; T. Padmanabhan, Phys. Rep. {\bf 406}, 49 (2005), [gr-qc/0311036].

\bibitem{Hai:2013ara} 
  H.~Hai, W.~Yong-Jiu and C.~Ju-Hua,
  Chin.\ Phys.\ B {\bf 22}, no. 7, 070401 (2013).

\bibitem{Futterman1988} J. A. Futterman, F. A. Handler, and R. A. Matzner,
{\it Scattering from black holes} (Cambridge University Press, England, 1988)
  
\bibitem{Matzner1977}  R. A. Matzner and M. P. Ryan, Phys. Rev. D {\bf 16}, 1636 (1977).

\bibitem {Westervelt1971} P. J. Westervelt, Phys. Rev. D {\bf 3}, 2319 (1971).

\bibitem{Peters1976} P. C. Peters, Phys. Rev. D {\bf 13}, 775 (1976).

\bibitem{Sanchez1976} N. G. S\'anchez, J. Math. Phys. {\bf 17}, 688 (1976);
N. G. S\'anchez, Phys. Rev. D {\bf 16} , 937 (1977);
N. G. S\'anchez, Phys. Rev. D {\bf 18}, 1030 (1978);
N. G. S\'anchez, Rev. D {\bf 18}, 1798 (1978).

\bibitem{Logi1977} W. K. de Logi and S. J. Kov\'acs, Phys. Rev. D {\bf 16}, 237 (1977).

\bibitem{Doram2002} C. J. L. Doran and A. N. Lasenby, Phys. Rev. D 66, 024006 (2002).

\bibitem{Dolan:2007ut} 
  S.~R.~Dolan,
  Phys.\ Rev.\ D {\bf 77}, 044004 (2008)
  doi:10.1103/PhysRevD.77.044004
  [arXiv:0710.4252 [gr-qc]].

\bibitem{Crispino:2009ki} 
  L.~C.~B.~Crispino, S.~R.~Dolan and E.~S.~Oliveira,
  Phys.\ Rev.\ D {\bf 79}, 064022 (2009)
  doi:10.1103/PhysRevD.79.064022
  [arXiv:0904.0999 [gr-qc]].


\bibitem{Churilov1974} A. A. Starobinsky and S. M. Churilov, Sov. Phys.- JETP {\bf 38}, 1 (1974).

\bibitem{Gibbons1975} G. W. Gibbons Commun. Math. Phys. {\bf 44}, 245 (1975)

\bibitem{Page1976} D. N. Page, Phys. Rev. D {\bf 13}, 198 (1976)




\bibitem{Churilov1973}  A. A. Starobinskii and S. M. Churilov, Zh. Eksp. Teor. Fiz. {\bf 65}, 3 (1973).

\bibitem{Crispino:2007zz} 
  L.~C.~B.~Crispino, E.~S.~Oliveira and G.~E.~A.~Matsas,
  Phys.\ Rev.\ D {\bf 76}, 107502 (2007).

\bibitem{Dolan:2009zza} 
  S.~R.~Dolan, E.~S.~Oliveira and L.~C.~B.~Crispino,
  Phys.\ Rev.\ D {\bf 79}, 064014 (2009)

\bibitem{Oliveira:2010zzb} 
  E.~S.~Oliveira, S.~R.~Dolan and L.~C.~B.~Crispino,
  Phys.\ Rev.\ D {\bf 81}, 124013 (2010).
 

 

\bibitem{Dolan}S. R. Dolan, E. S. Oliveira, L. C. B. Crispino, Phys. Lett. B {\bf 701}, 485 (2011).

\bibitem{ABP2012-1} M.~A.~Anacleto, F.~A.~Brito and E.~Passos, Phys. Rev. D {\bf 86}, 125015 (2012) 
[arXiv:1208.2615 [hep-th]]; Phys. Rev. D {\bf 87}, 125015 (2013) [arXiv:1210.7739 [hep-th]].

  
\bibitem{Anacleto:2015mta} 
  M.~A.~Anacleto, I.~G.~Salako, F.~A.~Brito and E.~Passos,
  Phys.\ Rev.\ D {\bf 92}, no. 12, 125010 (2015)
  doi:10.1103/PhysRevD.92.125010
  [arXiv:1506.03440 [hep-th]]; 
  M.~A.~Anacleto, F.~A.~Brito, A.~Mohammadi and E.~Passos,
  arXiv:1606.09231 [hep-th].  


\bibitem{Anacleto:2018acl} 
  M.~A.~Anacleto, F.~A.~Brito, J.~A.~V.~Campos and E.~Passos,
  arXiv:1810.13356 [hep-th].  
  
  
  
\bibitem{Brito2015} M.~A.~Anacleto, F.~A.~Brito and E.~Passos,
  Phys.\ Lett.\ B {\bf 743}, 184 (2015)
  [arXiv:1408.4481 [hep-th]].
  
\bibitem{Jung2004} E. Jung and D. Park, Class. Quantum Grav. {\bf 21}, 3717  (2004), arXiv:hep-th/0403251 [hep-th];
E. Jung, S. Kim, and D. Park, Phys. Lett. B {\bf 602}, 105 (2004), arXiv:hep-th/0409145 [hep-th].  

\bibitem{Doran2005} C. Doran, A. Lasenby, S. Dolan, and I. Hinder, Phys. Rev. D {\bf 71}, 124020 (2005), 
arXiv:gr-qc/0503019 [gr-qc].

\bibitem {Dolanprd2006} S. Dolan, C. Doran, and A. Lasenby, Phys. Rev. D {\bf 74}, 064005 (2006), arXiv:gr-qc/0605031 [gr-qc].

\bibitem{Castineiras2007} J. Castineiras, L. C. Crispino, and D. P. M. Filho, Phys. Rev. D {\bf 75}, 024012 (2007).
 
\bibitem{Benone:2014qaa} 
  C.~L.~Benone, E.~S.~de Oliveira, S.~R.~Dolan and L.~C.~B.~Crispino,
  Phys.\ Rev.\ D {\bf 89}, no. 10, 104053 (2014)
  doi:10.1103/PhysRevD.89.104053
  [arXiv:1404.0687 [gr-qc]].
  
\bibitem{Moura:2011rr} 
  F.~Moura,
  JHEP {\bf 1309}, 038 (2013)
  doi:10.1007/JHEP09(2013)038
  [arXiv:1105.5074 [hep-th]].

\bibitem{Paik:2017wcy} 
  S.~Gangopadhyay, R.~Mandal and B.~Paik,
  Int.\ J.\ Mod.\ Phys.\ A {\bf 33}, no. 14n15, 1850084 (2018)
  doi:10.1142/S0217751X18500847
  [arXiv:1703.10057 [gr-qc]].  

\bibitem{Kumar:2017hgs} 
  R.~Kumar and S.~G.~Ghosh,
  Eur.\ Phys.\ J.\ C {\bf 77}, no. 9, 577 (2017)
  doi:10.1140/epjc/s10052-017-5141-x
  [arXiv:1703.10479 [gr-qc]].
  
  
\bibitem{Rizzo:2006zb} 
  T.~G.~Rizzo,
  JHEP {\bf 0609}, 021 (2006)
  doi:10.1088/1126-6708/2006/09/021
  [hep-ph/0606051].
  
\bibitem{rani}  S. Rani, M.B. Amin and A. Jawad, 
  Eur. Phys. J. Plus {\bf131}, 436 (2016). doi.org/10.1140/epjp/i2016-16436-4
  
\bibitem{Mustafa:2019eet} 
  G.~Mustafa, G.~Abbas and T.~Xia,
  Chin.\ J.\ Phys.\  {\bf 60}, 362 (2019).
  doi:10.1016/j.cjph.2019.05.025
  
\bibitem{Marinho:2016ixt} 
  C.~I.~S.~Marinho and E.~S.~de Oliveira,
  arXiv:1612.05604 [gr-qc].

\bibitem{Anacleto:2017kmg} 
  M.~A.~Anacleto, F.~A.~Brito, S.~J.~S.~Ferreira and E.~Passos,
  Phys.\ Lett.\ B {\bf 788}, 231 (2019)
  doi:10.1016/j.physletb.2018.11.020
  [arXiv:1701.08147 [hep-th]].

\bibitem {Yennie1954} D. R. Yennie, D. G. Ravenhall, and R. N. Wilson, Phys. Rev. 95, 500 (1954).

\bibitem{Cotaescu:2014jca} 
  I.~I.~Cotaescu, C.~Crucean and C.~A.~Sporea,
  Eur.\ Phys.\ J.\ C {\bf 76}, no. 3, 102 (2016)
  doi:10.1140/epjc/s10052-016-3936-9
  [arXiv:1409.7201 [gr-qc]].
  
\bibitem{Das:1996we} 
  S.~R.~Das, G.~W.~Gibbons and S.~D.~Mathur,
  Phys.\ Rev.\ Lett.\  {\bf 78}, 417 (1997)
  doi:10.1103/PhysRevLett.78.417
  [hep-th/9609052]. 
     

\bibitem{Dolan:2012yc} 
  S.~R.~Dolan and E.~S.~Oliveira,
  Phys.\ Rev.\ D {\bf 87}, no. 12, 124038 (2013)
  doi:10.1103/PhysRevD.87.124038
  [arXiv:1211.3751 [gr-qc]].
  
\bibitem{Kim:2008vi} 
  W.~Kim, E.~J.~Son and M.~Yoon,
  JHEP {\bf 0804}, 042 (2008)
  doi:10.1088/1126-6708/2008/04/042
  [arXiv:0802.1757 [gr-qc]].  

\bibitem{Macedo:2016yyo} 
  C.~F.~B.~Macedo, L.~C.~B.~Crispino and E.~S.~de Oliveira,
  Int.\ J.\ Mod.\ Phys.\ D {\bf 25}, no. 09, 1641008 (2016)
  doi:10.1142/S021827181641008X
  [arXiv:1605.00123 [gr-qc]].

\bibitem{deOliveira:2018kcq} 
  E.~S.~de Oliveira,
  Eur.\ Phys.\ J.\ C {\bf 78}, no. 11, 876 (2018)
  doi:10.1140/epjc/s10052-018-6316-9
  [arXiv:1805.04987 [gr-qc]].
  
\end{thebibliography}
\end{document}